\documentclass[fleqn]{article}
\usepackage{amsmath,amsthm,amssymb,bm,lscape}
\usepackage[dvips]{graphicx}
\usepackage[body={17cm,25cm}]{geometry}

\newcommand{\Tr}{{\rm Tr}}

\def\Ths{{\mathsf \Theta}}

\def\Hs{{\mathsf H}}

\newcounter{app}
\newcommand{\app}[1]{
\refstepcounter{app}{\vspace{7mm}
\noindent\Large\bf Appendix
\theapp.
 \ #1 \par \vspace{5mm}}
\setcounter{equation}{0}
\def\theequation{\Alph{app}.\arabic{equation}}}
\begin{document}
\title{ New Q matrices and their functional equations  
for the eight vertex model at elliptic roots of unity}

\author{Klaus Fabricius
\footnote{e-mail Fabricius@theorie.physik.uni-wuppertal.de}\\
Physics Department, University of Wuppertal, 
42097 Wuppertal, Germany\\
Barry~M.~McCoy
\footnote{e-mail mccoy@max2.physics.sunysb.edu}\\               
C.N.Yang Institute for Theoretical Physics,\\ 
State University of New York,
 Stony Brook,  NY 11794-3840}
\date{\today}

\maketitle

\begin{abstract}

The Q matrix invented by Baxter in 1972 to solve the eight vertex
model at roots of unity exists for all values of $N$, the number of
sites in the chain, but only for a subset of roots of unity. We show in
this paper that a new Q matrix, which has recently been introduced and
is non zero only for $N$ even,  exists for all roots of unity. In addition
we consider the relations between
all of the known $Q$ matrices of the eight vertex model and conjecture
functional equations for them.


\end{abstract}

\section{Introduction}
In 1972 Baxter \cite{bax72} invented a method to compute the 
eigenvalues of the transfer matrix of the eight vertex model without
first computing the eigenvectors. This was done
by introducing an ``auxiliary''
matrix $Q(v)$ which satisfies the functional equation  
\begin{equation}
T(v)Q(v)=[h(v+\eta)]^NQ(v-2\eta)+[h(v-\eta)]^NQ(v+2\eta)
\label{tq}
\end{equation}
with
\begin{equation}
h(v)=\Ths_m(0)\Ths_m(-v)\Hs_m(v)
\label{hdef}
\end{equation}
where the quasiperiodic theta functions $\Ths_m(v),~\Hs_m(v)$ and the 
transfer matrix $T(v)$ of the eight vertex model 
are defined in appendix 1. The number
of lattice sites of the chain with periodic boundary conditions is $N$  
and $Q(v)$ satisfies the commutation relations
\begin{eqnarray}
&&[T(v),Q(v')]=0\label{tqcomm}\\
&&[Q(v),Q(v')]=0\label{qqcomm}
\end{eqnarray}

The equation (\ref{tq}) is obviously a matrix equation. However, the
commutation relations (\ref{tqcomm}) and (\ref{qqcomm}) allow all four
matrices in (\ref{tq}) to be simultaneously diagonalized and thus the
equation  also may be regarded as a scalar functional  equation for 
eigenvalues $t(v)$ and $q(v)$ of the matrices $T(v)$ and $Q(v)$. 

For any eigenvalue $t(v)$ the scalar tq equation may be considered 
to be a second order difference equation for $q(v)$. 
However, it is important to recognize that in addition to the scalar tq
equation quasi-periodicity properties must be independently specified for the
functions $q(v)$ in order to obtain explicit solutions  
and that, as explicitly demonstrated for the eight vertex model in
\cite{fm1}, the  solutions $q(v)$ to the scalar
tq equation do not have to satisfy the same quasi periodicity
conditions which are satisfied by the transfer matrix eigenvalues $t(v)$. This
difference in quasi-periodicity properties of $t(v)$ and $q(v)$ has
recently been studied in \cite{bm}. The importance of
this is that there are many models such as the SOS \cite{bax732} 
\cite{bax733} and RSOS \cite{abf} models for which the eigenvalues of the
transfer matrix  have been shown to satisfy the scalar tq equation but
an operator $Q(v)$ which satisfies a matrix TQ equation is not known.
It is therefore most interesting the ask the following question:

\vspace{.1in}

{\it What additional information is contained in a Q matrix which is not
contained in the scalar tq equation supplemented by the
quasiperiodicity properties of the eigenvalues $q(v)$?}   

\vspace{.1in}

This question is particularly relevant to the eight  vertex model where the
matrices constructed by Baxter in 1972 \cite{bax72} and in 1973
\cite{bax731} have been shown \cite{fm1} to be different. 
This lack of uniqueness occurs for the eight vertex model 
when the transfer matrix has degenerate
eigenvalues which occur when the parameter
$\eta$ satisfies the  ``root of unity'' condition imposed in the 1972
paper \cite{bax72}
\begin{equation}
2L_0\eta=2m_{10}K+im_{20}K'
\label{root}
\end{equation}
where $K~(K')$ are the complete elliptic integrals of the first kind
of modulus $k~(k')$ and $L_0,~m_{10}$ and $m_{20}$ are 
integers whose greatest common divisor is one.  
More generally the relation between quasiperiodicity and 
non-uniqueness of the
solutions to the scalar tq equation has been extensively investigated
by Bazhanov and Mangazeev \cite{bm} for the special case of 
$m_{20}=0$.

We have studied the non-uniqueness of $Q$ matrices for the eight vertex
model at various roots of unity (\ref{root}) in a series of papers 
\cite{fm1}-\cite{TQ} and in \cite{newQ} and \cite{TQ} we
have seen that there are cases of the root of unity condition
(\ref{root}) where by use of the methods of \cite{bax72} 
two different matrices may be constructed ,which we call $Q^{(1)}_{72}(v)$ and
$Q^{(2)}_{72}(v)$, that are distinct from the matrix $Q_{73}(v)$
constructed by Baxter \cite{bax731}. One of the distinguishing
features is that for different classes of the integers
$m_{10}$ and $m_{20}$ the three matrices may have different
commutation relations with the three discrete symmetry operators  
\begin{eqnarray}
&&S=\sigma^z\otimes\sigma^z\otimes \cdots \otimes \sigma^z\label{sdef}\\
&&R=\sigma^x\otimes\sigma^x\otimes \cdots \otimes \sigma^x\label{rdef}
\end{eqnarray}
and $RS=(-1)^NSR.$ 

A second most important property of $Q(v)$ matrices which goes beyond the
quasiperiodicity properties of the eigenvalues was presented in
\cite{fm1} where it was conjectured that the matrix $Q^{(1)}_{72}(v)$
satisfies a functional equation not involving $T(v)$. This equation is
specific to the specific matrix $Q^{(1)}_{72}(v)$  
and is NOT a consequence of the scalar tq equation and the
quasiperiodicity properties of the eigenvalues of $Q^{(1)}_{72}(v)$.
This functional equation is completely analogous to the functional
equation first found for the three state chiral Potts model
\cite{amp}. This analogue between the Q matrix of the eight vertex
model and the transfer matrix of the chiral Potts model is presented
in great generality in the 1990 paper of Baxter, Bazhanov and Perk
\cite{bbp}. However it is only the matrices $Q^{(1)}_{72}(v)$ and
$Q^{(2)}_{72}(v)$ for which this analogy will hold because no such
functional equation holds for $Q_{73}(v)$.

The purpose of this present paper is to extend the studies of
\cite{fm1}-\cite{TQ} in two ways. The first is to
demonstrate that the matrix $Q^{(2)}_{72}(v)$ which was studied in
\cite{TQ} for the case $m_{10}$ and $m_{20}$ both even may be extended
to all integer values of $m_{10}$ and $m_{20}$. The second is to exhibit
conjectured functional equations for all cases of the matrices
$Q^{(1)}_{72}(v)$ and $Q^{(2)}_{72}(v)$. In sec. 2 we formulate the
problem and summarize the results. The details of the construction of
$Q^{(2)}_{72}(v)$ for $m_{10}$ and $m_{20}$ not both even are given in
sec. 3. We conclude in sec. 4 with a discussion of our results and
a few open questions. 
 
\section{Formulation and summary of results}

The construction devised by Baxter in 1972 \cite{bax72} to find
matrices $Q(v)$ which satisfy (\ref{tq}) as summarized
in \cite{TQ} consists of three steps:

\vspace{.1in}
1) The construction of matrices $Q_R(v)$ and $Q_{L}(v)$ which satisfy
\begin{eqnarray}
&&T(v)Q_R(v)=\omega^{-N}[h(v+\eta)]^NQ_R(v-2\eta)
+\omega^{N}[h(v-\eta)]^NQ_R(v+2\eta)\label{tqr}\\
&&Q_L(v)T(v)=\omega^{-N}[h(v+\eta)]^NQ_L(v-2\eta)
+\omega^{N}[h(v-\eta)]^NQ_L(v+2\eta)\label{tql}
\end{eqnarray}
where $\omega$ is some phase (possibly equal to unity)
and $Q_{R,L}(v)$ are of the form
\begin{equation}
Q_{R,L}(v)_{\alpha,\beta} = \Tr[S_{R,L}(\alpha_{1},\beta_1)(v)
\cdots S_{R,L}(\alpha_N,\beta_N)(v)]
\label{TrS}
\end{equation}
where ${S_R}(\alpha_i,\beta_i)(v)$ are  matrices of some dimension
$L\times L$ and $\alpha_j,\beta_j=\pm.$
\vspace{.1in}

2) The establishing of interchange relations
\begin{equation}
Q_L(u)\Lambda Q_R(v)=Q_L(v)\Lambda Q_R(u)
\label{interchange}
\end{equation}
where there are four values of $\Lambda$ to be considered
\begin{equation}
\Lambda=I,S,R,RS=(-1)^NSR
\label{adef}
\end{equation}

3) The construction of $Q_{72}(v)$ from
\begin{equation}
Q_{72}(v)=Q_R(v)Q^{-1}_R(v_0)
\label{qdef}
\end{equation}
where $v_0$ is a value of the spectral parameter $v$ such that 
$Q_R(v_0)$ is nonsingular. Whenever  the interchange relation
(\ref{interchange}) holds for two different matrices $\Lambda _1$ 
and $\Lambda_2$
the matrix $Q_{72}(v)$ will satisfy
\begin{equation}
[Q_{72}(v),\Lambda_1\Lambda_2]=0
\label{qaacomm}
\end{equation}

\vspace{.1in}

The establishing of these three conditions is sufficient to prove
that the $Q(v)$ so defined will satisfy the commutation relations 
(\ref{tqcomm}) and (\ref{qqcomm}) and the $TQ$ equation 
(\ref{tq}) with the extra phase $\omega$.  If we set
\begin{equation}
Q_{72}(v)=\omega^{-Nv/2\eta}{\tilde Q}_{72}(v) 
\label{qtilde}
\end{equation}
then ${\tilde Q}_{72}(v)$ will satisfy (\ref{tq}) which has no phase factor
$\omega$ and it is obvious that ${\tilde Q}(v)$ will continue to
satisfy the commutation relations (\ref{tqcomm}) and (\ref{qqcomm}).

There are two choices for the matrices $S_{R,L}(\alpha, \beta)(v)$
which have been found to satisfy the requirements of steps 1-3. 
The first is the choice originally made by Baxter in 1972
\cite{bax72} where the only non zero elements are
$S^{(1)}_{R,L}(\alpha, \beta)_{k,k\pm 1}(v), 
S^{(1)}_{R,L}(\alpha, \beta)_{0,0}(v)$
and  $S^{(1)}_{R,L}(\alpha, \beta)_{L,L}(v)$ and the dimension $L$ is
the $L_0$ of (\ref{root}). This choice is valid for all $N$. 

The other choice, first found in \cite{newQ} is valid only for $N$ even
(because the trace  in (\ref{TrS}) vanishes identically for odd $N$). 
This choice is given for $k=1,\cdots L-1$ by
\begin{eqnarray}
&&{S_R}^{(2)}(+,\beta)_{k,k+1}(v) = -\Hs_m(v-t-2k\eta)\tau_{\beta,-k}
\label{SR1}\\
&&{S_R}^{(2)}(+,\beta)_{k+1,k}(v) =~~\Hs_m(v+t+2k\eta)\tau_{\beta,~k}
\label{SR2}\\
&&{S_R}^{(2)}(-,\beta)_{k,k+1}(v) = ~~\Ths_m(v-t-2k\eta)\tau_{\beta,-k}
\label{SR3}\\
&&{S_R}^{(2)}(-,\beta)_{k+1,k}(v) = ~~\Ths_m(v+t+2k\eta)\tau_{\beta,~k}
\label{SR4}
\end{eqnarray}
and
\begin{eqnarray}
&&{S_R}^{(2)}(+,\beta)_{1,L}(v)
=~~\Hs_m(v+t+2L\eta)\tau_{\beta,~L}\label{SR5}\\ 
&&{S_R}^{(2)}(+,\beta)_{L,1}(v)
=-\Hs_m(v-t-2L\eta)\tau_{\beta,-L}\label{SR6}\\
&&{S_R}^{(2)}(-,\beta)_{1,L}(v)
=~~\Ths_m(v+t+2L\eta)\tau_{\beta,~L}\label{SR7}\\
&&{S_R}^{(2)}(-,\beta)_{L,1}(v)
=~~\Ths_m(v-t-2L\eta)\tau_{\beta,-L}\label{SR8} 
\end{eqnarray}
and $S_L^{(2)}$ defined for $k=1,\cdots L-1$ by
\begin{eqnarray}
&&{S_L}^{(2)}(\alpha,+)_{k,k+1}(v) = ~~\Hs_m(v+t+2k\eta)\tau'_{\alpha,-k}
\label{SL1}\\
&&{S_L}^{(2)}(\alpha, +)_{k+1,k}(v) =-\Hs_m(v-t-2k\eta)\tau'_{\alpha,~k}
\label{SL2}\\
&&{S_L}^{(2)}(\alpha,-)_{k,k+1}(v) = ~~\Ths_m(v+t+2k\eta)\tau'_{\alpha,-k}
\label{SL3}\\
&&{S_L}^{(2)}(\alpha,-)_{k+1,k}(v) = ~~\Ths_m(v-t-2k\eta)\tau'_{\alpha,~k}
\label{SL4}
\end{eqnarray}
and
\begin{eqnarray}
&&{S_L}^{(2)}(\alpha,+)_{1,L}(v)
=-\Hs_m(v-t-2L\eta)\tau'_{\alpha,~L}\label{SL5}\\ 
&&{S_L}^{(2)}(\alpha,+)_{L,1}(v)
=~~\Hs_m(v+t+2L\eta)\tau'_{\alpha,-L}\label{SL6}\\
&&{S_L}^{(2)}(\alpha,-)_{1,L}(v)
=~~\Ths_m(v-t-2L\eta)\tau'_{\alpha,~L}\label{SL7}\\
&&{S_L}^{(2)}(\alpha,-)_{L,1}(v)
=~~\Ths_m(v+t+2L\eta)\tau'_{\alpha,-L}\label{SL8} 
\end{eqnarray}
where the dimension $L$ depends on $L_0$ and the parameters
$\tau_{\beta,k}$ and $\tau'_{\alpha,k}$ are arbitrary. The 
interchange relation 
(\ref{interchange}) will hold only when the parameter $t$ takes on
certain specific values. We note that
\begin{equation}
Q_L^{(2)}(v;t)=-Q^{(2)T}_R(2K-v;t)S
\label{qlqrt}
\end{equation}

We demonstrated in \cite{TQ} that there are three subcases of the root
of unity condition (\ref{root}) where $Q^{(1)}_{72}(v)$ satisfies
steps 1-3;
\begin{eqnarray}
{\rm case~I}~~~m_{10}~~{\rm odd}~~~m_{20}~~{\rm even}\\
{\rm case~II}~~~m_{10}~~{\rm odd}~~~m_{20}~~{\rm odd}\\
{\rm case~ III}~~m_{10}~~{\rm even}~~~m_{20}~~{\rm odd}
\end{eqnarray}
Furthermore in \cite{TQ} we demonstrated for case
of $m_{10}$ and $m_{20}$ both even where $Q^{(1)}_{72}(v)$ 
does not exist that $Q^{(2)}_{72}(v)$
does satisfy steps 1-3 for the two cases of $t=n\eta$ and $t=(n+1/2)\eta$
and in addition when $t=(n+1/2)\eta$ that there are four
subcases according to 
\begin{equation}
m_{10},m_{20}\equiv 0,2~({\rm mod}4)
\end{equation}

In this paper we show that the construction of $Q^{(2)}_{72}(v)$ of
\cite{TQ} with $t=n\eta$ may be extended to the cases $m_{10}$ and
$m_{20}$ not both even. To achieve this generalization we need to
allow the dimension $L$ of the matrices $S^{(2)}_{R,L}(\alpha,\beta)$
to be a multiple of the $L_0$ defined by (\ref{root}). Thus if we
rewrite (\ref{root}) as
\begin{equation}
2L\eta=2m_{1}K+im_{2}K'
\label{root2}
\end{equation}
where now $L,~m_1$ and $m_2$ are allowed to have common divisors.
We find that steps 1-3 are satisfied when $L,~m_1$ and $m_2$ are given
in terms of $L_0,~m_{10}$ and $m_{20}$ as shown in 
table \ref{tab:2}

\begin{table}[h!]
\center
\caption{Relation between the parameters occurring in (\ref{root}) 
and (\ref{root2}).}
\label{tab:2}
\begin{tabular}{|l|ll|lll|}\hline
&$m_{10}$&$m_{20}$&$m_{1}$&$m_{2}$&$L$\\ \hline
  I & odd&even&$2m_{10}$&$2m_{20}$&$2L_0$\\ 
  II & odd&odd&$4m_{10}$&$4m_{20}$&$4L_0$\\  
  III & even&odd&$4m_{10}$&$4m_{20}$&$4L_0$\\   \hline
\end{tabular}
\end{table}

As examples we have:

Case I: $\eta=K/3+iK'/3,~~m_1=2,m_2=4,L=6$

Case II: $\eta=K/3+iK'/6,~~m_1=4,m_2=4,L=12$

Case III:$\eta=K+iK'/4,~~m_1=8,m_2=4,L=8$

We summarize all of these results in tables \ref{tab:4}-\ref{tab:7} 
where we give the
matrices $\Lambda$ which satisfy the interchange relation
(\ref{interchange}) and indicate the cases where $Q_R(v)$ is nonsingular.

\begin{table}[h!]
\center
\caption{The interchange properties and the nonsingularity properties
  of $Q^{(1)}_{R}(v)$. We indicate by Y (or N)  that the interchange
  relation with $\Lambda$ holds (or fails). We indicate by Y (or N)
  that the inverse of $Q^{(1)}_R(v)$ exists (or fails to exist)}
\label{tab:4}
\begin{tabular}{|cc|cccc|c|}\hline
$m_{10}$&$m_{20}$&$I$&$S$&$R$&$RS$&$Q^{(1)-1}_R$\\\hline
odd&even&Y&Y&N&N&Y\\
odd&odd&N&Y&Y&N&Y\\
even&odd&Y&N&Y&N&Y\\
even&even&Y&Y&Y&Y&N\\\hline
\end{tabular}
\end{table}
\begin{table}[h!]
\center
\caption{The interchange properties 
  of $Q^{(2)}_{R}(v;n\eta)$ for $m_{10}$ and $m_{20}$ both even.
We indicate by Y (or N)  whether the interchange
  relation with $\Lambda$ holds (or fails) and the
  notation $0 (2)$ stands for $\equiv 0(2) ({\rm mod}4)$. In all cases
  the matrix $Q^{(2)}_R(v;n\eta)$ is nonsingular}
\label{tab:5}
\begin{tabular}{|cc|cccc|}\hline
$m_{10}$&$m_{20}$&$I$&$S$&$R$&$RS$\\\hline
0&0&Y&Y&N&N\\
2&0&Y&Y&N&N\\
0&2&Y&Y&N&N\\
2&2&Y&Y&N&N\\\hline
\end{tabular}
\end{table}

\begin{table}[h!]
\center
\caption{The interchange properties and nonsingularity properties
  of $Q^{(2)}_{R}(v;(n+1/2)\eta)$ for $m_{10}$ and $m_{20}$ both even.
We indicate by Y (or N)  that the interchange
  relation with $\Lambda$ holds (or fails). We indicate by Y (or N)
  that the inverse of $Q^{(2)}_R(v;(n+1/2)\eta)$ exists (or fails to exist).
The notation $0 (2)$ stands for $\equiv 0(2) ({\rm mod}4)$.}
\label{tab:6}
\begin{tabular}{|cc|cccc|c|}\hline
$m_{10}$&$m_{20}$&$I$&$S$&$R$&$RS$&$Q^{(2)-1}_R$\\\hline
0&0&Y&Y&N&N&Y\\
2&0&Y&Y&Y&Y&N\\
0&2&N&Y&Y&N&Y\\
2&2&Y&N&Y&N&Y\\\hline
\end{tabular}
\end{table}

\begin{table}[ht]
\center
\caption{The interchange properties 
  of  $Q^{(2)}_{R}(v;t)$ with $m_{10}$ and $m_{20}$ not both even. We indicate by Y (or N)  that the interchange
  relation with $\Lambda$ holds (or fails). We indicate by Y (or N)
  that the inverse of $Q^{(2)}_{R}(v;t)$ exists (or fails to exist). }
\label{tab:7}
\begin{tabular}{|cc|c|cc|c|c|cccc|c|} \hline
  $m_{10}$&$m_{20}$&$L_0$&$m_1$&$ m_2$&$L$&$t$&$I$&$S$&$R$&$RS$
&$Q^{(2)-1}_{R}(v;t)$\\ \hline
  o&e&e&$2m_{10}$&$2m_{20}$&$2L_0$&$2n\eta$&Y&Y&N&N&Y\\ \hline
  o&e&e&$2m_{10}$&$2m_{20}$&$2L_0$&$(2n+1)\eta$&Y&Y&Y&Y&N\\ \hline
  o&e&o&$2m_{10}$&$2m_{20}$&$2L_0$&$2n\eta$&Y&Y&Y&Y&N\\ \hline
  o&e&o&$2m_{10}$&$2m_{20}$&$2L_0$&$(2n+1)\eta$&Y&Y&N&N&Y\\ \hline
  o&o&e,o&$4m_{10}$&$4m_{20}$&$4L_0$&$n\eta$&Y&Y&Y&Y&Y\\ \hline
  e&o&e,o&$4m_{10}$&$4m_{20}$&$4L_0$&$n\eta$&Y&Y&Y&Y&Y\\ \hline
 \end{tabular}
\end{table}

\subsection{Quasiperiodicity of $Q^{(2)}_{72}(v;n\eta)$ for $m_{10}$
  and $m_{20}$ not both even}

The quasiperiodicity properties of 
$ Q^{(2)}_{72}(v;n\eta)$ 
are expressed in terms of  
\begin{equation}
\omega_1 = 2(r_1K+ir_2K') \hspace{0.3 in} \omega_2 = 2(bK+iaK') \hspace{0.3 in}
\label{om12def}
\end{equation}
where $r_1$ and $r_2$ are defined by
\begin{equation}
2m_{10}=r_0r_1,~~~~~m_{20}=r_0r_2
\label{defr}
\end{equation}
with $r_0$ the greatest common divisor in $2m_{10}$ and $m_{20}$ and
$a$ and $b$ are the integer solutions of
\begin{equation}
ar_1-br_2=1
\label{norm}
\end{equation}
We note the inverse relations
\begin{equation}
2K=a\omega_1-r_2\omega_2 \hspace{0.3in} 2iK'=-b\omega_1+r_1\omega_2
\label{inverseom}
\end{equation}
the relation
\begin{equation}
4L_0\eta=r_0\omega_1
\label{etaomega}
\end{equation}
and that for $m_{20}=0$ we have
\begin{equation}
r_0=2m_{10},~~r_1=a=1,~~r_2=b=0,~~\omega_1=2K,~~\omega_2=2iK'
\end{equation}

The following results are derived in appendix 3.

{\bf Case  I) For $m_{10}$ odd and $m_{20}$ even}
\begin{eqnarray}
&&Q_{72}^{(2)}(v+\omega_1;n\eta)
=SQ^{(2)}_{72}(v;n\eta)\label{eper1}\\
&& Q_{72}^{(2)}(v+\omega_2;n\eta)
=q'^{-N(1+r_2)}e^{-2\pi iNv/\omega_1}S^b
 Q^{(2)}_{72}(v;n\eta)\label{eper2}
\end{eqnarray}
where
\begin{equation}
q'=e^{i\pi \omega_2/\omega_1}\label{nomeqpdef}
\end{equation}
The area of the fundamental region $0,\omega_1,
\omega_1+\omega_2,\omega_2$ is $4KK'$.

If we note the definition (\ref{qtilde}) of ${\tilde Q}^{(2)}(v)$ we see
from (\ref{eper1}) and (\ref{eper2}) 
\begin{eqnarray}
&&{\tilde Q}_{72}^{(2)}(v+\omega_1;n\eta)
=S{\tilde Q}^{(2)}_{72}(v;n\eta)\label{teper1}\\
&& {\tilde Q}_{72}^{(2)}(v+\omega_2;n\eta)
=q'^{-N}e^{-2\pi iNv/\omega_1}S^b
 {\tilde Q}^{(2)}_{72}(v;n\eta)\label{teper2}
\end{eqnarray}
which  are identical with the
quasiperiodicity relations of $Q^{(1)}_{72}(v)$ for $N$ even which are 
reviewed in appendix 3.

{\bf Cases II and III) For $m_{20}$ odd}
\begin{eqnarray}
&&Q^{(2)}(v+\omega_1/2;n\eta)
=i^Ne^{i\pi N r_1r_2/4}RS^{r_1/2}Q^{(2)}(v;n\eta)
  \label{per1}\\
&&Q^{(2)}(v+\omega_2;n\eta)
=q'^{-N(1+r_2)}e^{-2\pi iNv/\omega_1}S^bQ^{(2)}(v;n\eta)\label{per2}
\end{eqnarray}
where we note that it follows from (\ref{defr}) that $r_0$ and $r_2$
must be odd and $r_1$ must be even. The size of the fundamental region
is $2KK'$. This is one half of the fundamental region of 
case I and is the size 
of the fundamental region of $Q_{73}(v).$

\subsection{ Degenerate eigenvalues of 
$Q^{(2)}_{72}(v;n\eta)$ for $m_{20}$ odd}

We have investigated the case $\eta=K/2+iK'/4~ (m_{10}=m_{20}=1)$ 
with $N=12$  and have discovered that $Q^{(2)}_{72}(v;n\eta)$
has 32 pairs of degenerate eigenvalues. The existence of degenerate
eigenvalues of a $Q(v)$ matrix is a new phenomenon not previously
seen. We take this to be evidence to support the following 

{\bf Conjecture}

\vspace{.1in}

1) The matrices $Q^{(2)}_{72}(v;n\eta)$ for $m_{20}$ odd always have
degenerate eigenvalues if $N$ is sufficiently large.
 
\vspace{.1in}

We also note from tables \ref{tab:4}-\ref{tab:7} that the only $Q_{72}(v)$
matrix which we have constructed using the procedure of Baxter's 1972
paper \cite{bax72} that shares the property  with the matrix
$Q_{73}(v)$ constructed in \cite{bax731} of commuting with all three
discrete symmetry operators $S,~R$ and $RS$ is $Q^{(2)}_{72}(v;n\eta)$ with
$m_{20}$ odd. However, $Q_{73}(v)$ and $Q^{(2)}_{72}(v;n\eta)$ for
$m_{20}$ odd are fundamentally different because  $Q_{73}(v)$  
has no degenerate eigenvalues.

\subsection{Bethe roots and L strings}

The eigenvalues $q(v)$ of any matrix $Q_{72}(v)$ are quasiperiodic
functions which may be characterized by the positions $v_j$ of their
zeros and from the scalar tq equation these positions satisfy the equation
\begin{equation}
0=h^N(v_j+\eta)q(v_j-2\eta)+h^N(v_j-\eta)q(v_j+2\eta)
\label{bethe1}
\end{equation}

There are two ways in which (\ref{bethe1}) can be satisfied.

I. If $q(v_j)=0$ and $q(v_j\pm 2\eta)\neq 0$ then we may write
(\ref{bethe1}) as 
\begin{equation}
\left(\frac{h(v_j+\eta)}{h(v_j-\eta)}\right)^N=
-\frac{q(v_j+2\eta)}{q(v_j-2\eta)}
\label{bethe}
\end{equation}
This equation is referred to as ``Bethe's equation'' and the $v_j$
must lie in the fundamental region of the quasiperiodic function
$h^N(v)$. We refer to these roots as Bethe roots and denote them at $v^B_j$. 

II.  In addition to these Bethe roots there may be sets containing $L$
roots $v_j$ of the form
\begin{equation}
v_{j;k}=v_{j;0}+2k\eta~~~0\leq k\leq L-1
\end{equation}
for which $q(v_{j;k})=q(v_{j,k}\pm 2\eta)=0$ and thus (\ref{bethe1}) is
identically satisfied for any $v_{j;0}$.
We refer to these sets of $L$ roots $v_j$ as $L$ strings. The
parameters $v_{j;0}$ will lie in the fundamental region of $Q_{72}(v)$.
These $L$ strings will cancel out from the scalar tq equation and as a
result the eigenvalues $t(v)$ of $T(v)$ are independent of $v_{j;0}$.

Each eigenvalue $q(v)$ of $Q_{72}(v)$ may be factorized as
\begin{equation}
q(v)=q_B(v)q_L(v)
\label{factor}
\end{equation}
where $q_B(v)$ contains the Bethe roots $v^B_j$ which 
are determined from (\ref{bethe}) and $q_L(v)$ contains the $L$ strings
whose centers $v_{j;0}$ are not determined from (\ref{bethe}). The
quasiperiodicity properties of $q_B(v)$  will in general vary from
eigenvalue to eigenvalue because the number of Bethe roots will in
general be different in each $q_B(v)$. The construction of
 \cite{bax732},\cite{bax733}, \cite{bax02} of 
a matrix whose eigenvalues are $q_B(v)$ relies 
on the explicit construction of the eigenvectors of $T(v)$ whose
eigenvalue $t(v)$ satisfies the scalar $tq$ equation with $q_B(v)$. This is
the opposite of the construction of either $Q_{72}(v)$ or $Q_{73}(v)$
which constructs a $Q(v)$ matrix without first computing the 
eigenvectors of $T(v)$.

The area of the fundamental region of all matrices $Q_{72}(v)$ 
studied in this paper except
for $Q_{72}^{(2)}(v;n\eta)$ with $m_{20}$ odd is determined from the
quasiperiodicity relations to be $4KK'$. However, the fundamental
region of $h(v)$ has the area $2KK'$. To see this we first note from 
(\ref{nHomega1})-(\ref{nTomega2}) that $h(v)$
has quasi periods $\omega_1$ and $\omega_2$. However it follows
further from (\ref{Thsh}) that for any
$s_B$ of the form
\begin{equation}
s_B={\rm even~integer}K+{\rm integer}~ iK'
\end{equation}
that 
\begin{equation}
h(v+s_B)=c_1e^{c_2v}h(v)
\end{equation}
where $c_1$ and $c_2$ are independent of $v$.
Thus if we write
\begin{eqnarray}
&&\omega_1/2=r_1K+r_2K'\\
&&\omega_2/2=bK+aiK'
\end{eqnarray}
we see that in cases where $r_1$ is even that $h(v)$ will have
quasiperiods $\omega_1/2,~\omega_2$ and when $r_1$ is odd that $b$ 
may be chosen even and thus
$h(v)$ will have quasiperiods $\omega_1,\omega_2/2$. 
The phase factor which is produced on the left hand side of
(\ref{bethe}) under the quasi periods $\omega_2$ or $\omega_2/2$ is
compensated for on the right hand side by the correct choice of the
parameter $\nu$ which occurs in the
exponential factor $e^{i\nu v}$ which is
present in the $q_B(v)$ of the factorization (\ref{factor}). Therefore in
these cases the area of the fundamental region of $h(v)$ is $2KK'$
which is one half of the area of the fundamental region of
$Q_{72}(v)$. If $r_1$ is even (odd) then $s_B$ is $\omega_1/2~(\omega_2/2)$.

\subsection{Functional equations for $Q^{(1)}_{72}(v)$ and 
${\tilde Q}^{(2)}_{72}(v)$}

In \cite{fm1} and \cite{fm2} we conjectured and verified in several
cases that for $m_{10}$ odd and $m_{20}=0$ the matrices $Q^{(1)}_{72}(v)$
satisfy the matrix functional equation
\begin{eqnarray}
& &\exp\left(-\frac{i\pi Nv}{2K}\right)Q^{(1)}_{72}(v-iK')\nonumber\\
&=&A\sum_{l=0}^{L-1}h^N(v-(2l+1)\eta)
\frac{Q^{(1)}_{72}(v)}{Q^{(1)}_{72}(v-2l\eta)Q^{(1)}_{72}(v-2(l+1)\eta)}
\label{con}
\end{eqnarray}
where $A$ commutes with $Q^{(1)}_{72}(v)$ and  is independent of $v$.
We therefore have investigated whether such a functional equation
will hold for $Q^{(1)}_{72}(v)$ and
${\tilde Q}^{(2)}_{72}(v)$ for all other values of $m_{20}$.
To make such an investigation we need to generalize
the shift in (\ref{con}) from $iK'$ to
\begin{equation}
s=s_1K+s_2iK'
\end{equation}
and adjust phase factors in (\ref{con}) to match the 
quasiperiodicity properties of
$Q_{72}^{(1)}(v)$ and ${\tilde Q}^{(2)}_{72}(v)$. Therefore 
we conjecture for all cases except ${\tilde Q}^{(2)}_{72}(v;n\eta)$ 
with $m_{20}$ odd that there is a value of $s$ such that
the matrix equation holds 
\begin{eqnarray}
& &\exp\left(-\frac{i\pi(-s_1r_2+s_2r_1)Nv}
{\omega_1}\right)Q_{72}(v-s)\nonumber\\
&=&A\sum_{l=0}^{L-1}h^N(v-(2l+1)\eta)
\frac{Q_{72}(v)}{Q_{72}(v-2l\eta)Q_{72}(v-2(l+1)\eta)}
\label{conq1}
\end{eqnarray}
where $Q_{72}(v)$ is either $Q^{(1)}_{72}(v)$ or ${\tilde Q}^{(2)}_{72}(v)$

We have determined the values of $s_1$ and $s_2$ by numerically
studying the conjecture in special cases and have found that the
functional equation (\ref{conq1}) holds for all the matrices
$Q^{(1)}_{72}(v)$, ${\tilde Q}_{72}^{(2)}(v;t)$ and 
${\tilde Q}^{(2)}(v;(n+1/2)\eta)$ with the single exception of
 ${\tilde Q}^{(2)}(v;n\eta)$ with $m_{20}$ odd where the matrix
shares with $Q_{73}(v)$ the property of commuting with all three
symmetry operators $S,~R$ and $RS$. We have also found that
there is no shift $s$ for which $Q_{73}(v)$ satisfies the functional
equation (\ref{conq1}).
The values of $s_1$ and
$s_2$ determined from these studies are given in tables 
\ref{tab:Q1}-\ref{tab:Q22}.

We also remark that if the shift $s$ is replaced by a shift $s'$ which
has the properties

\vspace{.1in}

1) $s'$ is a quasiperiod of $Q_{72}(v)$

2) the transformation $v\rightarrow v+s-s'=v+s_B$ leaves the 
   eigenvalues $q^o_B(v)$ invariant whose $N/2$ roots $v^B_j$ 
which lie in the fundamental region of $h^N(v)$
are determined by the Bethe's equation (\ref{bethe})
\vspace{.1in}

\noindent then for the eigenvalues $q^o_B(v)$ the 
matrix functional equation (\ref{conq1}) reduces to scalar functional
equations for the eigenvalues $q^o_B(v)$
\begin{equation}
A'\sum_{l=0}^{L-1}h^N(v-(2l+1)\eta)\frac{1}
{q^o_B(v-2l\eta)q^o_B(v-2(l+1)\eta)}=1
\label{funeqnon}
\end{equation}
with $A'$ is a constant whose value depends on the eigenvalue under 
consideration. 
The shifts $s'$ are given in tables \ref{tab:Q1}-\ref{tab:Q22}.

 \begin{table}[h!]
\center
\caption{Shifts $s$ and $s'$ for $Q^{(1)}_{72}(v)$. In the last column we
  indicate that discrete symmetry operator which commutes with
  $Q^{(1)}_{72}(v)$. The two shifts $s$ for $m_{10}$ and $m_{20}$ both odd
  are equivalent because they differ by a quasi-period.}
\label{tab:Q1}
\begin{tabular}{|c|c|c|c|c|c|}\hline
$m_{10}$&$m_{20}$&$s$&$s'$&quasiperiods&\\ \hline
odd&even&$iK'$&$2K$&$\omega_1,~\omega_2$&$S$\\
odd&odd&$iK',2K$&$2K+iK'$&$\omega_1/2,~2\omega_2$&$RS$\\
even&odd&$2K$&$iK'$&$\omega_1/2,~2\omega_2$&$R$\\\hline
\end{tabular}
\end{table}

\begin{table}[h!]
\center
\caption{Shifts $s$ and $s'$ for ${\tilde Q}^{(2)}(v;n\eta)$ for
  $m_{20}$ even. In all cases
$Q^{(2)}(v;n\eta)$ commutes with $S$.  There is no functional
  equation for $m_{20}$ odd}
\label{tab:Q21}
\begin{tabular}{|c|c|c|c|c|}\hline
$m_{10}$&$m_{20}$&$s$&$s'$&quasiperiods\\ \hline
odd&even&$iK'$&$2K$&$\omega_1,\omega_2$\\ 
even&even&$iK'$&$2K$&$\omega_1,\omega_2$\\ \hline
\end{tabular}
\end{table}

\begin{table}[h!]
\center
\caption{Shifts $s$ and $s'$ and quasiperiods 
for ${\tilde Q}^{(2)}(v;(n+1/2)\eta)$
for $m_{10}$ and $m_{20}$ both even.
The entries for $m_{10},m_{20}$ are $0(2)\equiv {\rm (mod~4)}$.
In the last column we
  indicate the discrete symmetry operator which commutes with
  ${\tilde Q}^{(2)}_{72}(v;(n+1/2)\eta)$. 
The two shifts $s$ for $m_{10}\equiv 0$ and $m_{20}\equiv 2$ (mod 4)   
are equivalent because they differ by a quasi-period.}
\label{tab:Q22}
\begin{tabular}{|c|c|c|c|c|c|}\hline
$m_{10}$&$m_{20}$&$s$&$s'$&quasiperiods&\\ \hline
0&0&$iK'$&$2K$&$\omega_1,~\omega_2$&$S$\\
0&2&$iK',2K$&$2K+iK'$&$\omega_1,~\omega_1/2+\omega_2$&$RS$\\
2&2&$2K$&$iK'$&$\omega_1,~\omega_1/2+\omega_2$&$R$\\\hline
\end{tabular}\end{table}

\subsection{Comparison of $Q^{(1)}_{72}(v)$ and  ${\tilde
    Q}^{(2)}_{72}(v;t)$ for $m_{10}$ odd and $m_{20}$ even}

We see from tables \ref{tab:Q1} and \ref{tab:Q21} that when $m_{10}$
is odd and $m_{20}$ is even the matrices  
$Q^{(1)}_{72}(v)$ and  ${\tilde  Q}^{(2)}_{72}(v;t)$ (with $t$ chosen
as in table\ref{tab:7})  
both satisfy the same functional
equation (\ref{conq1}). Furthermore we saw in section 2.1 that these
matrices satisfy the same quasiperiodicity equations
(\ref{teper1}),(\ref{teper2}).
It is therefore to be expected that the eigenvectors of these
matrices should be the same and that the ratios of the eigenvalues
should be independent of $v$. A numerical study of several special cases  
reveals that for $m_{10}$ odd and $m_{20}$ even the pair $Q^{(1)}_{72}(v)$ and
${\tilde Q}^{(2)}_{72}(v;t)$ 
in fact are similar up to proportionality
\begin{equation}
Q^{(1)}_{72}(v)={\rm const} M{\tilde
    Q}^{(2)}_{72}(v;t)M^{-1}~~~~m_{10}~{\rm odd},~m_{20}~{\rm even}\\
\end{equation}
We conjecture that this relation is generally true.

\subsection{Comparison of ${\tilde Q}^{(2)}_{72}(v;n\eta)$ 
and ${\tilde Q}^{(2)}(v;(n+1/2)\eta)$ for $m_{10},~m_{20}\equiv 0$ (mod 4)}
When $m_{10},~m_{20}\equiv 0$ 
(mod 4) we find from
tables \ref{tab:Q21} and \ref{tab:Q22} that   ${\tilde
  Q}^{(2)}_{72}(v;n\eta)$ and ${\tilde Q}^{(2)}(v;(n+1/2)\eta)$ also
satisfy (\ref{conq1}) with the same value of $s$. 
Furthermore  
the quasiperiodicity properties of   
${\tilde Q}^{(2)}_{72}(v,n\eta)$ 
and ${\tilde Q}^{(2)}(v;(n+1/2)\eta)$ are both given by
\begin{eqnarray}
&&{\tilde Q}^{(2)}_{72}(v+\omega_1)=S^{r_1}{\tilde Q}^{(2)}_{72}(v)\\
&&{\tilde Q}^{(2)}_{72}(v+\omega_2)=S^bq'^{-N}
e^{-2\pi iNv/\omega_1}{\tilde Q}^{(2)}_{72}(v)
\end{eqnarray}
and a numerical study of the case $L_{0}=3,~m_{10}=m_{20}=4~t=0,~\eta/2,~N=8$ 
reveals that 
${\tilde Q}_{72}(v;n\eta)$ and ${\tilde Q}^{(2)}(v;(n+1/2)\eta)$
in fact are similar. 
We conjecture that in general
\begin{equation}
{\tilde Q}^{(2)}_{72}(v;n\eta)
=M{\tilde  Q}^{(2)}_{72}(v;(n+1/2)\eta)M^{-1}
~~~m_{10},m_{20}\equiv 0~~({\rm mod}4)
\end{equation}

\section{Construction of $Q^{(2)}_{72}(v;t)$ for $m_{10}$ and $m_{20}$
  not both even}

We treat the steps 1-3 in separate subsections

\subsection{The equation for $TQ_R^{(2)}$ and  $Q_L^{(2)}T$ }
\label{TQ_R}

The study of $Q^{(2)}_{72}(v;t)$ for $m_{10}$ and $m_{20}$ not both even
closely parallels the study done in \cite{TQ} and wherever possible we
will refer to that paper for details of computations.
The principle
generalization needed is that for $m_{10}$ and $m_{20}$ both even
the dimension $L$ of the local matrices 
$S_R$ was $L=L_0$ where $L_0$ is determined from (\ref{root})
is odd and has no common divisors with $m_{10}$ and $m_{20}$.
In order to treat the cases where $m_{10}$ and $m_{20}$ are not both
even we will need to choose $L$ to be an even multiple of $L_0$ as
determined from (\ref{root}) and to define $m_1$ and $m_2$ as
(\ref{root2}).

We begin by following \cite{TQ} to show that 
when $Q^{(2)}_R(v;t)$ is determined from (\ref{SR1})-(\ref{SR8}) that 
(\ref{tqr}) with
\begin{equation}
\omega={\rm exp}\left(\frac{i\pi m_2}{2L}\right)
\label{2tq2}
\end{equation}
is valid for even L. This is (79) of \cite{TQ} and in appendix 3
of \cite{TQ} it is proven that for all  $L,m_1$ and $m_2$ related by
(\ref{root2}) that (\ref{tqr}) is valid if condition (C.31) of
\cite{TQ} holds
\begin{equation}
(\pm 1)^L\omega^L\frac{\Ths_m[(2L+1)\eta+t]}{\Ths_m(\eta+t)}=1
\label{sim4}
\end{equation} 
For even L equ. (\ref{sim4}) becomes
\begin{equation}
(-1)^{m_2/2} = 1
\end{equation}
and thus for (\ref{tqr}) to hold for even L 
we have to set $m_2 \equiv 0 ~({\rm mod}4)$.
We thus consider the three cases of 
table \ref{tab:2}

The companion matrix $Q^{(2)}_L(v)$ computed from (\ref{qlqrt}) 
satisfies (\ref{tql}) and  is given by (\ref{SL1})-(\ref{SL8}).

\subsection{The relation $Q_L(u)\Lambda Q_R(v)=Q_L(v)\Lambda Q_R(u)$.}

To proceed further we follow \cite{TQ} and
determine sufficient conditions for which the interchange relation
(\ref{interchange}). holds

The relation (\ref{interchange}) will hold if we can find similarity
transformation such that 
\begin{equation}
S^{(2)}_{L}(\alpha,\gamma)_{k,l}(u)\Lambda_{\gamma,\gamma'}
S^{(2)}_{R}(\gamma',\beta)_{k',l'}(v)=
Y_{k,k';m,m'}S^{(2)}_{L}(\alpha,\gamma)_{m,n}(v)
\Lambda_{\gamma,\gamma'}S^{(2)}_{R}(\gamma',\beta)_{m',n'}(u)
Y^{-1}_{n,n';l,l'} 
\label{YS}
\end{equation}
with diagonal matrix $Y$
\begin{equation}
Y_{k,k';m,m'}=y_{k,k'}\delta_{m,k}\delta_{k',m'}
\label{Ydiag}
\end{equation}
\begin{equation}
S^{(2)}_{L}(\alpha,\gamma)_{k,l}(u)\Lambda_{\gamma,\gamma'}
S^{(2)}_{R}(\gamma',\beta)_{k',l'}(v)=
\frac{y_{k,k'}}{y_{l,l'}}S^{(2)}_{L}(\alpha,\gamma)_{k,l}(v)
\Lambda_{\gamma,\gamma'}S^{(2)}_{R}(\gamma',\beta)_{k',l'}(u) 
\label{YS1}
\end{equation}
We write
\begin{equation}
S^{(2)}_{R}(\alpha,\beta)_{m,n}=\Phi^{\alpha}_{m,n}\tau^{\beta}_{m,n}
\hspace{0.6 in}
S^{(2)}_{L}(\alpha,\beta)_{m,n}=\tau^{'\alpha}_{m,n}\chi^{\beta}_{m,n}
\label{S_L}
\end{equation}
Then
\begin{equation}
S_{L}(\alpha,\gamma)_{k,l}\Lambda_{\gamma,\gamma'}(u)
S_{R}(\gamma',\beta)_{k'l'}(v) 
= \tau^{'\alpha}_{k,l}\chi^{\gamma}_{k,l}(u)
\Lambda_{\gamma,\gamma'}\Phi^{\gamma'}_{k',l'}(v)\tau^{\beta}_{k',l'}
\end{equation}
and thus (\ref{YS1}) is written for all four cases of $\Lambda$ as

\begin{equation}
\chi^{\gamma}_{k,l}(u)\Lambda_{\gamma,\gamma'}\Phi^{\gamma'}_{k',l'}(v)
=\frac{y_{k,k'}}{y_{l,l'}}\chi^{\gamma}_{k,l}(v)\Lambda_{\gamma,\gamma'}\Phi^{\gamma'}_{k',l'}(u)
\label{yeq}
\end{equation}

In sec. 6 of \cite{TQ} we explicitly evaluated (\ref{yeq}) 
for the case $m_{10}$ and $m_{20}$ both even. However, this evaluation
is also valid as well
for all the other cases of $m_{10}$ and $m_{20}$ and  we refer the reader to
\cite{TQ} for details of the computation. Thus, in what we hope 
is a more transparent notation, we find 
\begin{equation}
\frac{y_{k,k'}}{y_{k+1,k'+1}}=\frac{\hat{g}(u-v+2t+2(k+k')\eta)}{\hat{g}(v-u+2t+2(k+k')\eta)}
\label{ul_lr}
\end{equation}
\begin{equation}
\frac{y_{k,k'}}{y_{k+1,k'-1}}=\frac{\tilde{g}(u-v+2(k-k'+1)\eta)}{\tilde{g}(v-u+2(k-k'+1)\eta)}
\label{ur_ll}
\end{equation}
where the definitions of $\hat{g}, \tilde{g}$ is given in table \ref{tab:1}
 and the appendix.
\begin{table}[h!]
\center
\caption{Definition of $\hat{g}$ and $\tilde{g}$}
\label{tab:1}
\begin{tabular}{|c|c|c|c|c|}\hline
A&I&S&R&RS\\\hline
$\hat{g}$&$g^{-}_{\Ths\Ths}$&$g^{+}_{\Ths\Ths}$&$g^{-}_{\Hs\Ths}$&$g^{+}_{\Hs\Ths}$\\
\hline
$\tilde{g}$&$g^{+}_{\Ths\Ths}$&$g^{-}_{\Ths\Ths}$&$g^{+}_{\Hs\Ths}$&$g^{-}_{\Hs\Ths}$\\
\hline
\end{tabular}
\end{table}

The recursions (\ref{ul_lr}) and (\ref{ur_ll})
are interpreted as describing the transport on a torus of size
$L\times L$. Consequently in order to obtain a solution to these
equations we must show that from a set of initial values for $y_{k,l}$ all
remaining $y_{k,l}$ are determined consistently. This consistency
obtains only for certain values of $t$ and the cases of $L$ even and
odd must be treated separately.

 Consider first the case that $L$ is odd which was  treated in 
\cite{TQ}. The path shown in fig. 1 connects  an arbitrary point with 
its neighbor. It follows that all points on the torus can be 
reached by appropriate paths starting from a single point or
that e.g. all $y_{k,l}$ follow from $y_{1,1}$.

If $L$ is even we see from fig. 2 that there is no path connecting 
two neighboring points. The equations (\ref{ul_lr}) and (\ref{ur_ll}) 
thus form two disjoint sets. In this case all $y_{k,l}$
follow from two initial values e.g. $y_{1,1}$ and $y_{1,2}$.

These constructions of all $y_{k,l}$ from one or two initial values
will be consistent provided that transport of $y_{k,l}$ on a closed 
path has the result $y_{k,l}$.   
There are two cases to consider for $L$ odd and three cases for $L$ even.

For any $L$ we find for a closed path on the torus with winding 
numbers $(1,1)$ 
\begin{equation}
y_{k+L,k'+L}=y_{k,k'}\prod_{j=0}^{L-1}
\frac{\hat{g}(v-u+2t+2(k+k')\eta+4j\eta)}{\hat{g}(u-v+2t+2(k+k')\eta+4j\eta)}
\label{diag1}
\end{equation}
Similarly for any $L$ we find for a closed path for winding numbers $(1,-1)$
\begin{equation}
y_{k+L,k'-L}=y_{k,k'}\prod_{j=0}^{L-1}
\frac{\tilde{g}(u-v+2(k-k'+1)\eta+4j\eta)}{\tilde{g}(v-u+2(k-k'+1)\eta+4j\eta)}
\label{diag2}
\end{equation}

For even $L$ we have the additional condition that for the path : 
$(k,k')\rightarrow(k+1,k'+1)\rightarrow(k,k'+2)\cdots
(k,k'+L-2)\rightarrow(k+1,k'+L-1)\rightarrow(k,k'+L)$ shown in Fig. 2
which has winding numbers $(0,1)$
we obtain from equs. (\ref{ul_lr}) and (\ref{ur_ll})
\begin{equation}
y_{k,k'+L}=y_{k,k'}\prod_{j=0}^{L/2-1}
\frac{\hat{g}(u-v+2t+2(k+k')\eta+4j\eta)\tilde{g}(u-v+2(k-k'+1)\eta+4j\eta)}
     {\hat{g}(v-u+2t+2(k+k')\eta+4j\eta)\tilde{g}(v-u+2(k-k'+1)\eta+4j\eta)}
\label{hor}
\end{equation}
Thus for $L$ odd there are two conditions for the existence of a solution 
of (\ref{ul_lr}) and (\ref{ur_ll}) \begin{equation}
y_{k+L,k'+L}=y_{k,k'} \hspace{0.3 in}
y_{k+L,k'-L}=y_{k,k'}
\end{equation}
while for $L$ even there are three conditions
\begin{equation}
y_{k,k'+L}  =y_{k,k'} \hspace{0.3 in}
y_{k+L,k'+L}=y_{k,k'} \hspace{0.3 in}
y_{k+L,k'-L}=y_{k,k'}
\end{equation}

The case of odd $L$ was considered in \cite{TQ}. Here we consider the
case of $L$ even with $m_1$ and $m_2$ even as indicated in table
\ref{tab:2}. Then using the (anti) symmetry properties
(\ref{f0mm}),(\ref{f1pm}),(\ref{g0mm}),(\ref{g1pm})
we find that (\ref{diag1}) and (\ref{diag2}) become for all $\Lambda$
\begin{equation}
\prod_{j=0}^{L-1}
\frac{\hat{g}(v-u+2t+2(k+k')\eta+4j\eta)}{\hat{g}(v-u-2t-2(k+k')\eta-4j\eta)} = 1
\label{diag11}
\end{equation}
\begin{equation}
\prod_{j=0}^{L-1}
\frac{\tilde{g}(v-u+2(k-k'+1)\eta+4j\eta)}{\tilde{g}(v-u-2(k-k'+1)\eta-4j\eta)} = 1
\label{diag21}
\end{equation}
and that (\ref{hor}) becomes
\begin{equation}
\prod_{j=0}^{L/2-1}
\frac{\hat{g}(u-v+2t+2(k+k')\eta+4j\eta)\tilde{g}(u-v+2(k-k'+1)\eta+4j\eta)}
     {\hat{g}(u-v-2t-2(k+k')\eta-4j\eta)\tilde{g}(u-v-2(k-k'+1)\eta-4j\eta)}
\label{hor1} = \pm 1
\end{equation} 
where for $L/2=$ even the right hand side is $+1$ for $\Lambda=I,S,R,RS$ and
for $L/2=$ odd  the right hand side is $+1$ for $\Lambda=I,S$ and $-1$ 
for $\Lambda=R,RS$

To determine the consistency of (\ref{diag11})-(\ref{hor1}) we need
the periodicity properties for even $m_1$ and $m_2 \equiv 0 (\rm mod
4)$ which follow from (\ref{f0mper}),(\ref{f1pper}),(\ref{g0mper}),
(\ref{g1pper}) of the appendix
\begin{equation}
 g^{-}_{\Ths\Ths}(u+2L\eta)=g^{-}_{\Ths\Ths}(u)
\hspace{0.85 in}
g^{+}_{\Ths\Ths}(u+2L\eta)=g^{+}_{\Ths\Ths}(u)
\label{periodTT}
\end{equation} 
\begin{equation}
g^{-}_{\Hs\Ths}(u+2L\eta)=(-1)^{m_1/2}g^{-}_{\Hs\Ths}(u)
\hspace{0.3 in}
g^{+}_{\Hs\Ths}(u+2L\eta)=(-1)^{m_1/2}g^{+}_{\Hs\Ths}(u)
\label{periodHT}
\end{equation} 

We will treat the cases of $\Lambda=I,S$ and $\Lambda=R,RS$ separately

\subsubsection{The cases $\Lambda=I,S$.}
\label{AS} . 
We set 
\begin{equation}
t=n\eta
\label{teta}
\end{equation}
and will use the periodicity conditions (\ref{periodTT}) to prove 
that in each of the  equs. (\ref{diag11}), (\ref{diag21}),
(\ref{hor1})  for each factor
$\hat{g}$ and  $\tilde{g}$ in the numerator there exist values of $n$
such that there is a 
corresponding factor in the denominator with the same 
argument modulo the period $4L\eta$ for (\ref{diag11}), (\ref{diag21})
or $2L\eta$ for (\ref{hor1}). 
 
Consider first (\ref{diag11}). The difference of 
the argument of the $j$
factor in the numerator with the $j'$ factor in the denominator is 
\begin{equation}
{\rm diff_1(j,j')} = 4(n+k+k'+j+j')\eta = \frac{n+k+k'+j+j'}{L} 4L\eta
\end{equation}
Similarly for equ. (\ref{diag21}) the corresponding difference is
\begin{equation}
{\rm diff_2(j,j')} = 4(k-k'+1+j+j')\eta = \frac{k-k'+1+j+j'}{L} 4L\eta
\end{equation}
We see that for fixed integer $0\leq n,k,k',j<L-1$ there is an integer 
$j'$ with $0\leq j'<L$ such that 
${\rm diff_i(j,j')}$ with $i=1,2$ is an integer multiple of $4L\eta$.
This proves that when $t$ is given by (\ref{teta}) with $n$  an
integer the  equations.  (\ref{diag11}), (\ref{diag21}) are satisfied.

We finally consider the more restrictive equ. (\ref{hor1}).
The difference of arguments of factors $\hat{g}$ are
\begin{equation}
{\rm diff_{h1}(j,j')} = 4(n+k+k'+j+j')\eta = \frac{n+k+k'+j+j'}{L/2} 2L\eta
\end{equation}
and the difference of arguments of functions $\tilde{g}$ is
\begin{equation}
{\rm diff_{h2}(j,j')} = 4(k-k'+1+j+j')\eta = \frac{k-k'+1+j+j'}{L/2} 2L\eta
\end{equation}
written in a form accommodated to the shorter range $0 \leq j' < L/2$.
As the period of $\hat{g}$ and $\tilde{g}$ is 
$2L\eta$ as shown in equ. (\ref{periodTT})
we see that when $n$ is an integer that (\ref{hor1}) is satisfied.
We conclude that when $t$ is given by (\ref{teta}) with $n$ integer
the interchange relation (\ref{interchange}) for $\Lambda=I$ and $\Lambda=S$ 
is satisfied in all cases  listed in table \ref{tab:2}.  
It thus follows from (\ref{qaacomm}) that $Q^{(2)}(v)$ commutes with $S$.

\subsubsection{The cases $\Lambda=R,RS$}

The interchange relation (\ref{interchange})
for  cases $\Lambda=R,RS$ is examined using the (anti)periodicity conditions
(\ref{periodHT}). The proof of (\ref{diag11}) and (\ref{diag21})
given above for $\Lambda=I,S$ required the periodicity of $4L\eta$ 
and thus the identical proof works for $\Lambda=R,RS$. However
the proof of (\ref{hor1}) depends on whether $m_1/2$ and $L/2$ are
even or odd and these cases will be treated separately.

{\bf Cases II and III of table \ref{tab:2}}

We see from table \ref{tab:2} that $m_1/2$ is even for the cases II and III
and thus it follows from (\ref{periodHT}) that 
$g^{\pm}_{\Hs\Ths}(u)$ have the period $2L\eta$. 
We also see from table \ref{tab:2} in cases II and III that $L/2$ is even and 
thus the right hand side of equ. (\ref{hor1}) is $+1$
Therefore the proof given in subsection \ref{AS} works also in this
case.
We conclude that for cases II and III the interchange relation
(\ref{interchange})
holds for $A=R$ when $t$ is given by (\ref{teta}) with $n$ integer.
Thus from (\ref{qaacomm}) $Q^{(2)}(v;n\eta)$ commutes with $R$ as well as
with $S$ (and hence also with $RS$) 

{\bf Case I of table \ref{tab:2} with  $L/2=L_0$ even}
\label{oddeven}

We see from table \ref{tab:2} that $m_1/2=m_{10}$ is odd
for case I and thus  it follows from  (\ref{periodHT}) that 
$g^{\pm}_{\Hs\Ths}(u)$ are antiperiodic with the period $2L\eta$.
When  $L/2=L_0$ is even the right hand side of (\ref{hor1}) is $+1$.

To find values of $n$ such that  (\ref{hor1}) will hold  in this case 
consider 
$\hat{g}(u-v+2n\eta+2(k+k')\eta+4j\eta)$ in the numerator and 
 $\hat{g}(u-v-2n\eta-2(k+k')\eta-4j'\eta)$ in the denominator. 
The difference of their arguments is 
\begin{equation}
{\rm diff_1(j,j')} = 4(n+k+k'+j+j')\eta = \frac{n+k+k'+j+j'}{L/2} 2L\eta
\end{equation}
As $L/2$ is an integer then for any integer $n$ 
the difference ${\rm diff_1(j,j')}$ may become 
${\hat m} 2L\eta$, where ${\hat m}$ is an integer. 
If ${\hat m}$=even the respective
functions drop out of the product on the left hand side of
(\ref{hor1}). If ${\hat m}$=odd the two functions drop out up to a minus sign.
If for a pair $j,j'$ with $j'\neq j$ the functions drop out there is another pair of functions $j',j$ which also drop out.
It follows that if $j'\neq j$ two pairs will drop out without sign change.
So we have only to inspect the case $j'=j$.

Similarly consider a function $\tilde{g}(u-v+2(k-k'+1)\eta+4j\eta)$ 
in the numerator and a function
$\tilde{g}(u-v-2(k-k'+1)\eta-4j'\eta)$ in the denominator. 
The difference of their arguments is 
\begin{equation}
{\rm diff_2(j,j')} = 4(k-k'+1+j+j')\eta = \frac{k-k'+1+j+j'}{L/2} 2L\eta
\end{equation}
Thus for integer $n$ the difference  ${\rm diff_2(j,j')}$ 
may become ${\tilde m} 2L\eta$, where ${\tilde m}$ is an integer. 
If ${\tilde m}$=even 
the respective functions drop out of the product on the left hand side
of (\ref{hor1}). If ${\tilde m}$=odd the two functions drop out up 
to a minus sign.Thus as before
it follows that if $j'\neq j$ two pairs will drop out without sign change.
So we have only to inspect the case $j'=j$.

When $j=j'$ and $n$ even then because $L/2$ is even  we see that 
if $k+k'$ is even (odd) there are two (zero) solutions 
$j$ with  $0\leq j < L/2$ of
\begin{equation}
 \frac{n+k+k'+2j}{L/2} = {\rm integer} 
\end{equation}
and if there are two solutions one of these produces a factor $-1$ and 
the other a factor $+1$.
Similarly  if $k+k'$ (and thus $k-k'$ is even (odd)) 
there are zero (two) solutions $j$ of \begin{equation}
 \frac{k-k'+1+2j}{L/2} =  {\rm integer} 
\end{equation}
and  if there are two solutions these produce a factor $-1$ and 
the other a factor $+1$.
This results in a factor $-1$ after all functions on the 
left hand side of (\ref{hor1}) have dropped out and
it follows that for $m_1/2$=odd and $L/2$=even the interchange
relation is not satisfied for $A=R$ and $A=RS$ if $t=2l\eta$. 

In the opposite case $j=j'$ and $n$ odd
Then for $k+k'$ even (odd) there are zero (two) or two solutions 
$j$ with $ 0\leq j < L/2$ of
\begin{equation}
 \frac{n+k+k'+2j}{L/2} =  {\rm integer}
\end{equation}
and zero (two) or two solutions $j$ of
\begin{equation}
 \frac{k-k'+1+2j}{L/2} =  {\rm integer}
\end{equation}
Thus all factors $-1$ appear in pairs.
It follows that for $m_1/2$=odd and $L/2$=even the interchange 
relation is satisfied for $A=R$ and $A=RS$ if $t=(2l+1)\eta$.

{\bf Case I of table \ref{tab:2} with $L/2=L_0$  odd.}

In this case the right hand side of equ. (\ref{hor1}) is $-1$. 
This is the only difference with the case
$m_1/2$ odd and $L/2$ even. It follows that the results of 
section \ref{oddeven} are reversed.
Thus for $t=(2l+1)\eta$ the interchange relation 
is valid for $A=I,S$ and for $t=2l\eta$   
the interchange relation is valid for $A=I,S,R,RS$

\subsection{The matrix $Q^{(2)}(v;n\eta)$}

It remains to compute $Q^{(2)}(v;t)$ from $Q^{(2)}_R(v)$  by using
\begin{equation}
Q^{(2)}(v;t)=Q_R^{(2)}(v)Q_R^{(2)-1}(v_0)
\end{equation}
and for this construction to be valid the matrix $Q^{(2)}_R(v)$ must be
non singular for some value of $v$. While no analytic results are
available we have investigated this question numerically for 
examples
of all three cases of table \ref{tab:2} for systems of size
$N=8$. The conclusions of this study are given in table \ref{tab:7}

>From the validity of the interchange relation for all $\Lambda$
it follows for $m_{20}$ odd that
\begin{equation}
[Q^{(2)}(v),S]=[Q^{(2)}(v),R]=[Q^{(2)}(v),RS]=0
\label{caseIIcomm}
\end{equation}
which are the same symmetry properties of the transfer matrix $T(v)$

For $m_{10}$ odd and $m_{20}$ even and
the choice of $t$ given in table \ref{tab:7} 
 the commutation properties $Q^{(2)}(v;t)$ are 
\begin{equation}
[Q^{(2)}(v;t),S]=0,~~ [Q^{(2)}(v;t),R]\neq 0,
~~ [Q^{(2)}(v;t),RS]\neq 0
\label{caseIcomm}
\end{equation}

\section{Discussion and open questions}

The studies of Q matrices, beginning with \cite{bax72},\cite{bax731}
and continuing through \cite{fm1}-\cite{TQ} have revealed that the
concept of a Q matrix is not unique and that for a full understanding
it is necessary to study several essentially different constructions.
For example the construction  1973  \cite{bax731} exists for generic $\eta$
for an even number of sites and commutes with both symmetry operators
$S$ and $R$ but is of limited use in determining the degeneracy of the
transfer matrix eigenvalues at roots of unity. The Q matrix defined in
the 1972 paper \cite{bax72} is defined for all N and, because it fails
to commute with the operator $R$, is very useful in characterizing the
degeneracies of the transfer matrix but does not exist when 
$m_{10}$ and $m_{20}$ are both even \cite{TQ}. 
For this excluded case new Q matrices, 
which exist only for $N$ even, were found in \cite{newQ},
\cite{TQ} and \cite{roan} and these new Q matrices are shown
in \cite{newQ} and \cite{TQ} to reveal the full degeneracy of the
transfer matrix eigenvalues.

Particularly for the most important case of real $\eta$ 
it seems somewhat misleading and unnatural that different forms of Q
should be used for different classes of roots of unity. In this paper
we have overcome this dichotomy for even N by demonstrating that 
the new Q matrix of \cite{newQ} and \cite{TQ} exists for all
roots of unity. Furthermore
we have extended the functional equation for
$Q_{72}(v)$ first conjectured in \cite{fm1} for the special case
$m_{20}=0$ and $m_{10}$ odd to all values of $m_{10}$ and $m_{20}$
where a matrix $Q_{72}(v)$ constructed by the method of \cite{fm1}
exists.

However, there are several open questions which
have been raised that need further investigation:

1) In both \cite{TQ} and the present paper we have found the
   commutation of the Q matrices with the discrete symmetry operators
   depends on the parity of the numbers $m_{10}$ and $m_{20}$.  
   This property has not been anticipated in the literature and needs
   further explanation. 
  
2) The discovery in this paper that for some cases Q matrices 
   constructed by the
   method of \cite{bax72} can have degenerate eigenvalues is totally
   unexpected.

3)  For the case $m_{20}=0$ and $m_{10}$ even 
it was observed in \cite{roan} that the
$Q_{72}(v)$ matrix is intimately connected with the construction
used by Baxter in \cite{bax731}-\cite{bax733} to obtain the
SOS models and the eigenvectors of the eight vertex model. Furthermore 
in \cite{roan} this connection is exploited to prove the functional
equation for $Q$. It is thus most interesting to see whether this
connection with \cite{bax731}-\cite{bax733} extends to all the matrices 
$Q^{(2)}_{72}(v;t)$ considered in this paper. This is particularly
interesting in the case $m_{20}=0$ and $m_{10}$ odd where
${\tilde Q}^{(2)}_{72}(v;n\eta)$ is similar up to proportionality 
to $Q^{(1)}_{72}(v)$ because $Q^{(1)}_{72}(v)$ is defined for all $N$.
and not just $N$ even.

4) The  matrix of 1973 \cite{bax731} which exists for generic $\eta$ 
and  the new Q matrix of
   \cite{newQ}, \cite{TQ} and \cite{roan} which is shown in this  present 
paper to exist for all roots of unity are only defined for  $N$ even. 
The matrix of 1972 \cite{bax72}
   exists for all N but only when $m_{10}$ and $m_{20}$ are not both
   even. Consequently there is as yet NO Q matrix for the case of
   $m_{10}$ and $m_{20}$ both even and $N$ odd. This is perhaps the
   most interesting and challenging of all the cases of the eight vertex
   model \cite{fm2},\cite{bax89}-\cite{baz06}

\app{The functions $\Hs_m(v),\Ths_m(v)$ and the transfer matrix $T(v)$}

The standard definition of the theta functions $\Hs(v)$ and $\Ths(v)$
are
\begin{eqnarray}
 &&\Hs(v)=2\sum_{n=1}^{\infty}(-1)^{n-1}q^{(n-\frac{1}{2})^2}
\sin[(2n-1)\pi v/(2K)]\\
&&\Ths(v)=1+2\sum_{n=1}^{\infty}(-1)^nq^{n^2}\cos(nv\pi/K)
\end{eqnarray}
where
\begin{equation}
q=e^{-\pi K'/K}
\label{nomeqdef}
\end{equation}
In this paper we will use the ''modified'' theta functions $\Hs_m(v)$
and $\Ths_m(v)$ of \cite{bax731} defined by 
\begin{equation}
\Hs_m(u) = \exp\left(\frac{i\pi m_{20}(u-K)^2}{8KL_0\eta}\right)\Hs(u)
\hspace{0.4 in}
\Ths_m(u) = \exp\left(\frac{i\pi m_{20}(u-K)^2}{8KL_0\eta}\right)\Ths(u)
\end{equation}
The functions $\Hs_m(v)$ and $\Ths_m(v)$ are themselves theta functions
\cite{TQ} with (quasi)periods $\omega_{1,2}$ given in  
in (28)-(31) \cite{TQ} by
\begin{eqnarray}
&&\Hs_m(v+\omega_1) = (-1)^{r_1+r_1r_2}\Hs_m(v)
\label{Homega1}\\
&&\Ths_m(v+\omega_1) =(-1)^{r_1r_2}\Ths_m(v)
\label{Tomega1}\\
&&\Hs_m(v+\omega_2) 
= (-1)^{b+ab}q'^{-1}\exp\left(-\frac{2\pi i(v-K)}{\omega_1}\right)\Hs_m(v)
= (-1)^{a+b+ab}q'^{-(1+r_2)}\exp\left(-\frac{2\pi iv}{\omega_1}\right)\Hs_m(v)
\nonumber\\
\label{Homega2}\\
&&\Ths_m(v+\omega_2) = 
(-1)^{ab}q'^{-1}\exp\left(-\frac{2\pi i(v-K)}{\omega_1}\right)\Ths_m(v)
=(-1)^{a+ab}q'^{-(1+r_2)}\exp\left(-\frac{2\pi iv}{\omega_1}\right)\Ths_m(v)
\nonumber\\
\label{Tomega2}
\end{eqnarray}
We can slightly simplify these relations if we note firstly from the 
definition (\ref{norm}) that a and b cannot both be even and thus
$a+b+ab$ must be odd. Thus 
\begin{equation}
(-1)^{a+b+ab}=-1~~~{\rm and}~~~(-1)^{a+ab}=(-1)^{1+b}
\label{abrel}
\end{equation}
Furthermore, since by definition $(r_1,r_2)=1$ the identical argument
shows that
\begin{equation}
(-1)^{r_1+r_2+r_1r_2}=-1~~~{\rm and}~~~(-1)^{r_1+r_1r_2}=(-1)^{1+r_2}
\label{r12rel}
\end{equation}
Thus we will use (\ref{Homega1})-(\ref{Tomega2}) in the slightly 
simpler form \begin{eqnarray}
&&\Hs_m(v+\omega_1) = (-1)^{1+r_2}\Hs_m(v)
\label{nHomega1}\\
&&\Ths_m(v+\omega_1) =(-1)^{r_1r_2}\Ths_m(v)
\label{nTomega1}\\
&&\Hs_m(v+\omega_2) 
= -q'^{-(1+r_2)}\exp\left(-\frac{2\pi iv}{\omega_1}\right)\Hs_m(v)
\label{nHomega2}\\
&&\Ths_m(v+\omega_2) 
=(-1)^{1+b}q'^{-(1+r_2)}\exp\left(-\frac{2\pi iv}{\omega_1}\right)\Ths_m(v)
\label{nTomega2}
\end{eqnarray}
 We also recall (A.8)-(A.13) of \cite{TQ} 
\begin{equation}
H_m(2K-v) = H_m(v),
\hspace{0.5 in}
\Theta_m(2K-v) = \Theta_m(v)
\label{help1}
\end{equation}
\begin{equation}
H_m(-v) = -\exp\left(\frac{i\pi m_2 v}{2L\eta}\right)H_m(v)
\hspace{0.5 in}
\Theta_m(-v) = \exp\left(\frac{i\pi m_2 v}{2L\eta}\right)\Theta_m(v)
\label{Hminus}
\end{equation}
\begin{equation}
\Theta_m(v+iK') 
= i q^{-1/4}\exp\left(\frac{-i\pi m_1v}{2L\eta}\right)C
H_m(v),~~~ H_m(v+iK') 
= i q^{-1/4}\exp\left(\frac{-i\pi m_1v}{2L\eta}\right)C\Theta_m(v)
\label{Thsh}
\end{equation}
where
\begin{equation}
C = \exp\left(\frac{\pi m_2K'}{8KL\eta}(2K-iK')\right)
\label{ThshC}
\end{equation}
\begin{equation}
H_m(u+2L_0\eta) = (-1)^{m_{10}}i^{m_{10}m_{20}}
\left\{\begin{array}{ll}
H_m(u) & \mbox{ if $m_{20}$=even}\\
\Theta_m(u) & \mbox{ if $m_{20}$=odd}\\
\end{array}\right. \
\label{Hm2Leta}
\end{equation}
\begin{equation}
\Theta_m(u+2L_0\eta) = i^{m_{10}m_{20}}
\left\{\begin{array}{ll}
\Theta_m(u) & \mbox{ if $m_{20}$=even}\\
\Hs_{m}(u) & \mbox{ if $m_{20}$=odd}\\
\end{array}\right. \
\label{Tm2Leta}
\end{equation}

For odd $m_{20}$ the integers $r_0$ and $r_2$ are odd and $r_1$ is even. Thus 
\begin{equation}
2L_0\eta = \omega_1/2+\omega_1(r_0-1)/2
\end{equation}
with $(r_0-1)/2$  integer
and using (\ref{defr}), (\ref{nHomega1}) and (\ref{nTomega1}) we write
(\ref{Hm2Leta}) and (\ref{Tm2Leta}) as
\begin{eqnarray}
\label{Hper2} 
&&\Hs_m(v+\frac{\omega_1}{2}) 
= (-1)^{r_1/2}\exp(\frac{i\pi r_1r_2}{4})\Ths_m(v)\\ 
&&\Ths_m(v+\frac{\omega_1}{2}) =\exp(\frac{i\pi r_1r_2}{4})\Hs_m(v) \nonumber
\end{eqnarray}

The modified theta functions have the following identities:
\begin{eqnarray}
&&\Ths_m(u)\Ths_m(v)+\Hs_m(u)\Hs_m(v) = 
\frac{2q^{1/4}}{\Hs(K)\Ths(K)}\exp\left(\frac{i\pi m_{20}K'^{2}}{8KL\eta}\right) \nonumber \\
&&\Hs_m((u+v+iK')/2)\Hs_m((u+v-iK')/2)\Hs_m((iK'+u-v)/2+K)\Hs_m((iK'-u+v)/2+K)
\end{eqnarray}
\begin{eqnarray}
&&\Ths_m(u)\Ths_m(v)-\Hs_m(u)\Hs_m(v) = 
\frac{2q^{1/4}}{\Hs(K)\Ths(K)}\exp\left(\frac{i\pi m_{20}(K'^{2}-4K^{2})}{8KL\eta}\right) \nonumber \\
&&\Hs_m((iK'+u-v)/2)\Hs_m((iK'-u+v)/2)\Hs_m((u+v+iK')/2+K)\Hs_m((u+v-iK')/2-K)
\end{eqnarray}
\begin{eqnarray}
&&\Ths_m(u)\Hs_m(v)+\Hs_m(u)\Ths_m(v) = 
\frac{2}{\Hs(K)\Ths(K)} \nonumber \\
&&\Hs_m((u+v)/2)\Ths_m((u+v)/2)\Hs_m((u-v)/2+K)\Ths_m((u-v)/2+K)
\end{eqnarray}
\begin{eqnarray}
&&\Hs_m(u)\Ths_m(v)-\Ths_m(u)\Hs_m(v) = 
\frac{2}{\Hs(K)\Ths(K)}\exp\left(\frac{-i\pi m_{20}K}{2L\eta}\right) \nonumber \\
&&\Hs_m((u-v)/2)\Ths_m(-(u-v)/2)\Hs_m((u+v)/2+K)\Ths_m((u+v)/2-K)
\end{eqnarray}

The transfer matrix of the eight vertex model is \cite{bax731}
\begin{equation}
T(v)|_{\alpha,\beta}={\rm Tr}W(\alpha_1,\beta_1)W(\alpha_2,\beta_2)\cdots 
W_(\alpha_N,\beta_N)
\end{equation}
where the Boltzmann weights $W(\alpha,\beta)$ are $2\times 2$ matrices
with the non zero matrix elements given by
\begin{eqnarray}
&&W(1,1)|_{1,1}=W(-1,-1)|_{-1,-1}=\Ths_m(-2\eta)\Ths_m(\eta-v)\Hs_m(\eta+v)\\
&&W(-1,-1)|_{1,1}=W(1,1)|_{-1,-1}=-\Ths_m(-2\eta)\Hs_m(\eta-v)\Ths_m(\eta+v)\\
&&W(-1,1)|_{1,-1}=W(1,-1)|_{-1,1}=-\Hs_m(-2\eta)\Ths_m(\eta-v)\Ths_m(\eta+v)\\
&&W(1,-1)|_{1,-1}=W(-1,1)|_{-1,1}=\Hs_m(-2\eta)\Hs_m(\eta-v)\Hs_m(\eta+v)
\end{eqnarray}

\app{The functions $f^{\pm}_{\Ths\Ths}(u), f^{\pm}_{\Hs\Ths}(u),
g^{\pm}_{\Ths\Ths}(u), g^{\pm}_{\Hs\Ths}(u)$}

We recall the definitions of \cite{TQ}
\begin{equation}
f^{+}_{\Ths\Ths}(u) = ~~\Hs_m((u+iK')/2)\Hs_m((u-iK')/2)
\hspace{0.2 in}
g^{+}_{\Ths\Ths}(u) = ~~\Hs_m((iK'+u)/2+K)\Hs_m((iK'-u)/2+K)
\end{equation}
\begin{equation}
g^{-}_{\Ths\Ths}(u) = -\Hs_m((iK'+u)/2)\Hs_m((iK'-u)/2)
\hspace{0.2 in}
f^{-}_{\Ths\Ths}(u) = -\Hs_m((iK'+u)/2+K)\Hs_m((u-iK')/2-K)
\end{equation}
\begin{equation}
f^{+}_{\Hs\Ths}(u) = \Hs_m(u/2)\Ths_m(~~u/2)
\hspace{0.2 in}
g^{+}_{\Hs\Ths}(u) = \Hs_m(u/2+K)\Ths_m(u/2+K)
\end{equation}
\begin{equation}
g^{-}_{\Hs\Ths}(u) = \Hs_m(u/2)\Ths_m(-u/2)
\hspace{0.2 in}
f^{-}_{\Hs\Ths}(u) = \Hs_m(u/2+K)\Ths_m(u/2-K)
\end{equation}
\begin{equation}
\Ths_m(u)\Ths_m(v)+\Hs_m(u)\Hs_m(v) = \frac{2q^{1/4}}{\Hs_1(0)\Ths_1(0)}
\exp\left(\frac{i\pi m_2K'^{2}}{8KL\eta}\right)f^{+}_{\Ths\Ths}(u+v)g^{+}_{\Ths\Ths}(u-v)
\end{equation}
\begin{equation}
\Ths_m(u)\Ths_m(v)-\Hs_m(u)\Hs_m(v) = 
\frac{2q^{1/4}}{\Hs_1(0)\Ths_1(0)}\exp\left(\frac{i\pi m_2(K'^{2}-4K^{2})}{8KL\eta}\right)
g^{-}_{\Ths\Ths}(u-v)f^{-}_{\Ths\Ths}(u+v)
\end{equation}
\begin{equation}
\Ths_m(u)\Hs_m(v)+\Hs_m(u)\Ths_m(v) = \frac{2}{\Hs_1(0)\Ths_1(0)}f^{+}_{\Hs\Ths}(u+v)g^{+}_{\Hs\Ths}(u-v)
\end{equation}
\begin{equation}
\Hs_m(u)\Ths_m(v)-\Ths_m(u)\Hs_m(v) =\frac{2}{\Hs_1(0)\Ths_1(0)} 
\exp\left(\frac{-i\pi m_2K}{2L\eta}\right)g^{-}_{\Hs\Ths}(u-v)f^{-}_{\Hs\Ths}(u+v)
\end{equation}

For $m_1$ and $m_2$ both even it is proven in \cite{TQ} that
\begin{equation}
g^{-}_{\Ths\Ths}(u+2L\eta) = (-1)^{m_{2}/2}(-1)^{m_1m_2/4}g^{-}_{\Ths\Ths}(u)
\label{f0mper}
\end{equation}
\begin{equation}
g^{-}_{\Ths\Ths}(-u) = g^{-}_{\Ths\Ths}(u)
\label{f0mm}
\end{equation}
\begin{equation}
g^{+}_{\Ths\Ths}(u+2L\eta) = (-1)^{m_1m_2/4}g^{+}_{\Ths\Ths}(u)
\label{f1pper}
\end{equation}
\begin{equation}
g^{+}_{\Ths\Ths}(-u) = g^{+}_{\Ths\Ths}(u)
\label{f1pm}
\end{equation}
\begin{equation}
g^{-}_{\Hs\Ths}(u+2L\eta) = (-1)^{(m_1+m_2)/2}(-1)^{m_1m_2/4}g^{-}_{\Hs\Ths}(u)
\label{g0mper}
\end{equation}
\begin{equation}
g^{-}_{\Hs\Ths}(-u) =-g^{-}_{\Hs\Ths}(u)
\label{g0mm}
\end{equation}
\begin{equation}
g^{+}_{\Hs\Ths}(u+2L\eta) = (-1)^{m_1/2}(-1)^{m_1m_2/4}g^{+}_{\Hs\Ths}(u)
\label{g1pper}
\end{equation}
\begin{equation}
g^{+}_{\Hs\Ths}(-u) = g^{+}_{\Hs\Ths}(u)
\label{g1pm} 
\end{equation}

\app{Quasiperiodicity} 

We here derive the quasiperiodicity relations  
(\ref{eper1}),(\ref{eper2}),(\ref{per1}),(\ref{per2}) 
for $Q^{(2)}(v;n\eta)$ for $m_{10}$ and $m_{20}$ not both even and
review several of the computations in \cite{TQ}. 

\vspace{.1in}

{\bf $Q^{(2)}_{72}(v;n\eta)$ for $m_{10}$ odd and $m_{20}$ even}

The quasiperiodicity properties of $Q^{(2)}_{72}(v;n\eta)$ for
$m_{10}$ odd are derived
in an identical fashion to the case with $m_{10}$ even in \cite{TQ}.In
both cases $r_0$ is even and we find from (246) and (250) of \cite{TQ}
(and the last line of (\ref{Homega2}) and (\ref{Tomega2})) that
\begin{equation}
S_R^{(2)}(\alpha,\beta)(v+\omega_1)
=(-\alpha)^{r_1}(-1)^{r_1r_2}S_R^{(2)}(\alpha,\beta)(v)
\end{equation}
and
\begin{equation}
S^{(2)}_R(\alpha,\beta)(v+\omega_2)=(-1)^{nr_0/2}\alpha^b(-1)^{a+b+ab}
q'^{-(1+r_2)}
e^{-2\pi iv/\omega_1}MS_R(\alpha,\beta)(v)M^{-1}
\end{equation}
with
\begin{equation}
M_{k,k'}=\delta_{k,k'}e^{-\pi i r_0k(k-1)/(2L)}
(-1)^{nr_0k/2}e^{-\pi inr_0k/(2L)}
\end{equation}
Thus using (\ref{sdef}) and (\ref{TrS}) (and the fact that $N$ is
even)
we find for $m_{10}$ either even or odd
\begin{eqnarray}
&&Q^{(2)}_{72}(v+\omega_1;n\eta)=S^{r_1}Q_{72}^{(2)}(v;n\eta)
\label{help2}\\
&&Q^{(2)}_{72}(v+\omega_2:n\eta)
=q'^{-N(1+r_2)}e^{-2\pi iNv/\omega_1}S^bQ(v;n\eta)
\end{eqnarray}
In the case where $m_{10}$ is odd it follows from (\ref{defr}) that
$r_1$ is odd and (\ref{help2}) reduces to (\ref{eper1}).

\vspace{.1in}

{\bf $Q^{(2)}_{72}(v;n\eta)$ for $m_{20}$ odd}

In this case we recall that $r_0$ and $r_2$ are odd and $r_1$ is even.

{\bf Proof of relation (\ref{per1}).}

\vspace{.1in}

For convenience we set 
\begin{equation}
C = \exp\left(\frac{i\pi r_{1}r_{2}}{4}\right)
\label{C}
\end{equation}
and find from (\ref{Hper2})  that

\begin{equation}
{S^{(2)}_R}(\alpha,\beta)_{k,k+1}(v+\frac{\omega_1}{2}) = 
 iC(\sigma_1\sigma_3^{r_{1}/2})_{\alpha,\gamma}(~~i{S^{(2)}_R}(\gamma,\beta)_{k,k+1}(v))
\label{SR1q}
\end{equation}
\begin{equation}
{S^{(2)}_R}(\alpha,\beta)_{k+1,k}(v+\frac{\omega_1}{2}) =  
iC(\sigma_1\sigma_3^{r_{1}/2})_{\alpha,\gamma}(-i{S^{(2)}_R}(\gamma,\beta)_{k,k+1}(v))
\label{SR2q}
\end{equation}
We perform a similarity transformation to remove the 
factors $\pm i$ in front of $S^{(2)}_R$ on the right
hand side.
\begin{equation}
{\tilde{S^{(2)}_R}}(\alpha,\beta)_{k,l}(v)=A_{kk'}{S^{(2)}_R}(\alpha,\beta)_{k'l'}(v)A^{-1}_{l',l}
\label{sim}
\end{equation}
where $A$ is a diagonal $L\times L$ matrix.
\begin{equation}
A_{kl}=a_k\delta_{kl}
\label{A}
\end{equation}
\begin{equation}
{\tilde{S}^{(2)}_R}(\alpha,\beta)_{k,k+1}(v+\frac{\omega_1}{2})=
iC(\sigma_1\sigma_3^{r_{1}})_{\alpha,\gamma}(~~i\frac{a_k}{a_{k+1}}{S^{(2)}_R}(\gamma,\beta)_{k,k+1}(v))
\label{St}
\end{equation}
\begin{equation}
{\tilde{S_R}^{(2)}}(\alpha,\beta)_{L,1}(v+\frac{\omega_1}{2})~~~= 
iC(\sigma_1\sigma_3^{r_{1}})_{\alpha,\gamma}(~~i\frac{a_L}{a_{1}}{S^{(2)}_R}(\gamma,\beta)_{k,k+1}(v))
\label{StL}
\end{equation}
In (\ref{St}) we set
\begin{equation}
a_{k+1} = ia_k   
\label{ak}
\end{equation}
from which it follows that
\begin{equation}
a_k = i^{k-1}a_1
\end{equation}
and thus in (\ref{StL})
\begin{equation}
i\frac{a_L}{a_{1}} =  i^{L} =1
\label{akL}
\end{equation}
because  $L=4L_0$. Thus we find
\begin{equation}
{\tilde{S}^{(2)}_R}(\alpha,\beta)_{k,l}(v+\frac{\omega_1}{2})=
iC(\sigma_1\sigma_3^{r_{1}/2})_{\alpha,\gamma}
{S^{(2)}_R}(\gamma,\beta)_{k,l}(v)
\label{St1}
\end{equation}
which gives (\ref{per1}) when inserted into (\ref{TrS}).

\vspace{.1in}

{\bf Proof of relation (\ref{per2}).}

\vspace{.1in}
We find from (\ref{nHomega2}) and (\ref{nTomega2}) that 
with $C(v)=\exp(-2\pi i v/\omega_1)$
\begin{equation}
{S_R}(\alpha,\beta)_{k,k+1}(v+\omega_2) = 
-q'^{-(1+r_2)}C(v) \exp(~~\frac{4\pi ik\eta}{\omega_1})\exp(~~\frac{2\pi it}{\omega_1}) (\sigma_3)^{b}_{\alpha,\gamma}{S_R}(\gamma,\beta)_{k,k+1}(v)
\label{SR1p3}
\end{equation}
\begin{equation}
{S_R}(\alpha,\beta)_{k+1,k}(v+\omega_2) = 
-q'^{-(1+r_2)}C(v)\exp(-\frac{4\pi ik\eta}{\omega_1})\exp(-\frac{2\pi it}{\omega_1})(\sigma_3)^{b}_{\alpha,\gamma}{S_R}(\gamma,\beta)_{k+1,k}(v)
\label{SR2p3}
\end{equation}
We perform a similarity transformation to remove the expressions $\exp(\pm 4\pi ik\eta/\omega_1)\exp(\pm\frac{2\pi it}{\omega_1})$ on the right
hand sides.
\begin{equation}
{\tilde{S}_R}(\alpha,\beta)_{k,l}(v)=A_{kk'}{S_R}(\alpha,\beta)_{k'l'}(v)A^{-1}_{l',l}
\label{sim2}
\end{equation}
where $A$ is a diagonal $L\times L$ matrix.
\begin{equation}
A_{kl}=a_k\delta_{kl}
\label{A1}
\end{equation}
\begin{equation}
a_{k+1}=\exp\left(\frac{4\pi ik\eta}{\omega_1}\right)\exp\left(\frac{2\pi it}{\omega_1}\right) a_k
\label{A2}
\end{equation}
It follows that
\begin{equation}
a_{k}=\exp\left(\frac{2\pi i\eta k(k-1)}{\omega_1}\right)\exp\left(\frac{2\pi it(k-1)}{\omega_1}\right) a_1
\label{A3}
\end{equation}
Which removes the expressions $\exp(\pm 4\pi ik\eta/\omega_1)\exp(\pm 2\pi it/\omega_1)$ in (\ref{SR1p3}) and (\ref{SR2p3}) for $k<L$.
For $k=L$ we need that
\begin{equation}
\frac{a_L}{a_1}\exp(4\pi iL\eta/\omega_1)\exp(2\pi it\eta/\omega_1) =1
\label{A4}
\end{equation}
must be satisfied. We set $t = n\eta$. Then
\begin{equation}
\frac{a_L}{a_1}\exp(4\pi iL\eta/\omega_1)\exp(2\pi in\eta/\omega_1) = \exp(2\pi ir_0L(L+1)/(4L_0))\exp(2\pi inr_0L/(4L_0))
\label{A5}
\end{equation}
Thus, noting from table\ref{tab:5}  for $m_{20}$ odd that $L=4L_0$ we
see that (\ref{A4}) holds. Therefore we have shown that 
\begin{equation}
{\tilde{S}_R}(\alpha,\beta)_{k,l}(v+\omega_2) = 
-q'^{-(1+r_2)}\exp\left(-\frac{2\pi iv}{\omega_1}\right){\sigma_3^{b}}_{\alpha,\gamma}{S_R}(\gamma,\beta)_{k,l}(v)
\label{SR1px}
\end{equation}
which gives (\ref{per2}) when used in (\ref{TrS}).\\
{\bf Quasiperiodicity for $Q^{(1)}_{72}(v)$ for $m_{10}$ odd and
    $m_{20}$ even}

It follows from (\ref{defr}) when $m_{10}$ is odd and $m_{20}$ is even
that $r_0$ is even and $r_1$ is odd. Therefore (271) of \cite{TQ}
becomes
\begin{equation}
S^{(1)}_R(\alpha, \beta)(v+\omega_1)
=(-1)^{1+r_2}{\sigma_3}_{\alpha,\gamma} S^{(1)}_{R}(\gamma,\beta)
\end{equation}
and using (\ref{sdef}) and (\ref{TrS}) we obtain
\begin{equation}
Q^{(1)}_{72oe}(v+\omega_1)=(-1)^{N(1+r_2)}SQ^{(1)}_{72oe}(v)
\end{equation}

>From (220) of \cite{TQ} and (\ref{abrel}) we have
\begin{equation}
S_R^{(1)}(\alpha,\beta)(v+\omega_2)=(-1)^{(1+a)}q'^{-1}
\exp\left(-\frac{2\pi iv}{\omega_1}\right)M^{(1)}{\sigma_3}^b_{\alpha,\gamma}S^{(1)}_R(\gamma,\beta)(v)M^{(1)-1}
\label{final}
\end{equation}
with
\begin{equation}
M^{(1)}_{k,k'}=e^{-2\pi i\eta k(k-1)/\omega_1}\delta_{k,k'}
\end{equation}
Inserting (\ref{final}) into (\ref{TrS}) and using (\ref{sdef}) we obtain
\begin{equation}
Q^{(1)}_{72}(v+\omega_2)=(-1)^{N(1+a)}q'^{-N}
\exp\left(-\frac{2\pi iNv}{\omega_1}\right)S^bQ^{(1)}_{72}(v)
\end{equation}

\newpage
\begin{figure}[tp]
\centering  
\includegraphics[angle=0,width=12cm]{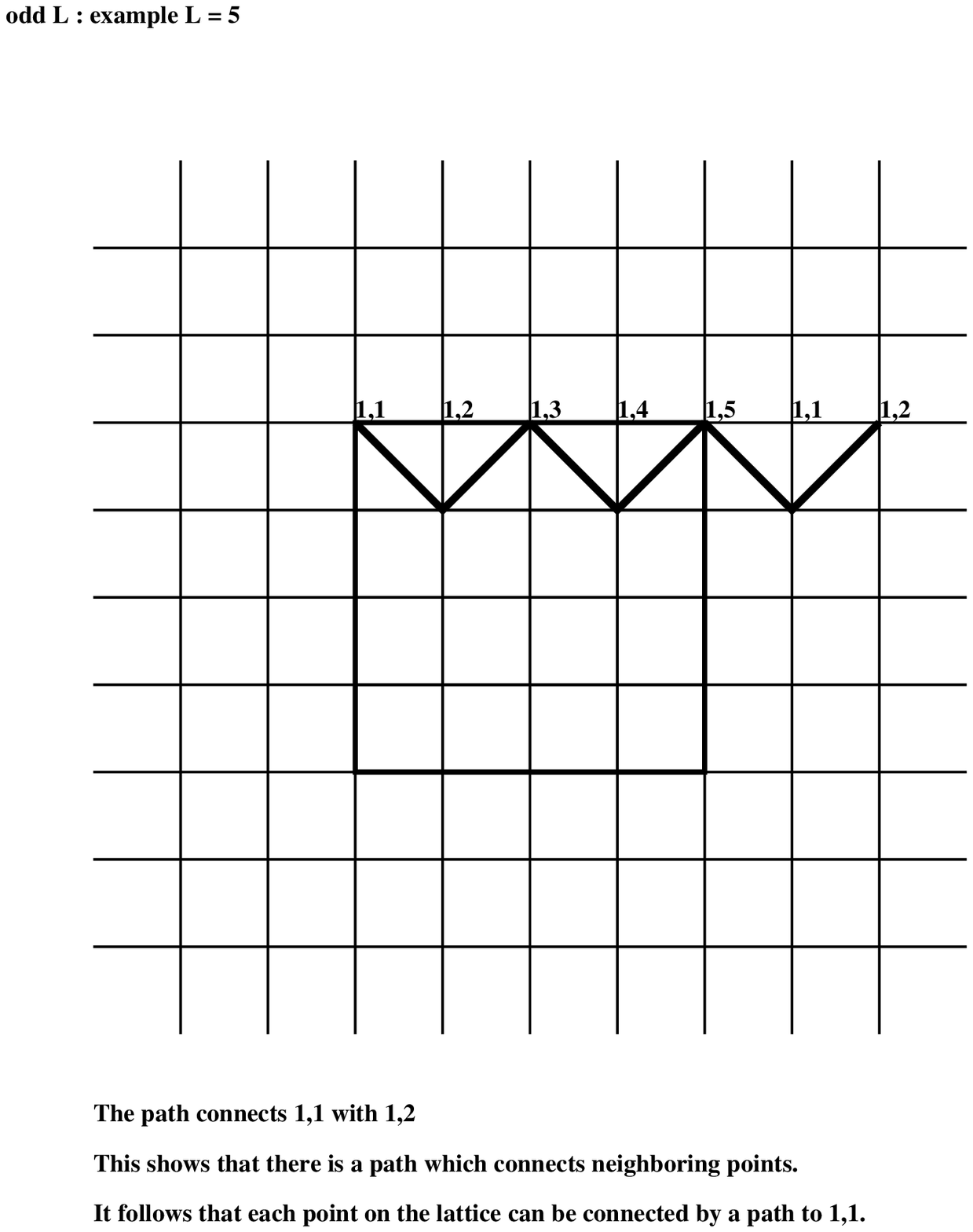}
\caption{Paths.} 
\label{fig1}
\end{figure} 
\newpage
\begin{figure}[tp]
\centering  
\includegraphics[angle=0,width=12cm]{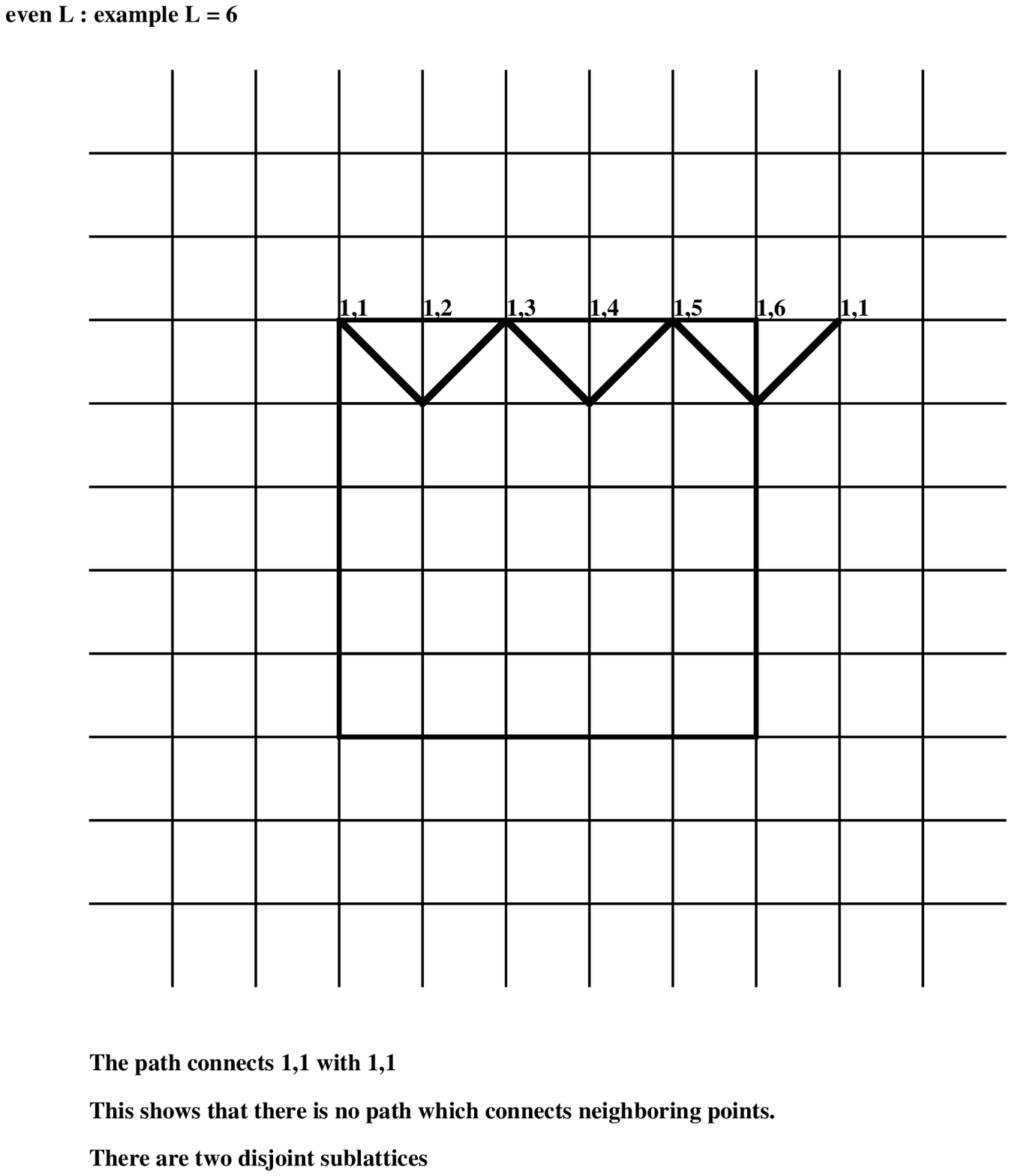}
\caption{Paths.} 
\label{fig2}
\end{figure} 

\end{document}